\documentstyle[preprint,aps]{revtex}
\tightenlines
\begin{document}
\draft
\title{The onset of a liquid-vapour transition in metallic nanoparticles}

\author{ G. Bilalbegovi\'c and H. O. Lutz}

\address{Fakult\"at f\"ur Physik, Universit\"at Bielefeld,
D-33615 Bielefeld, Germany}

\date{\today}

\maketitle

\begin{abstract}

We study lead nanodroplets containing $147$ to $1415$ 
atoms at temperatures ranging from the bulk melting point 
up to the beginning of the evaporation regime. 
Molecular dynamics simulation 
and an embedded atom potential are used. The structures,
total energies, and mobility of atoms in the clusters are analyzed.
We found that the liquid cluster of $147$ atoms shows a pronounced
tendency to form non-spherical shapes, and sometimes separates
into two droplets. Bigger clusters 
disassemble by evaporation of monomers. 
We also explore shape oscillations of these nanodroplets using 
the nuclear liquid drop model.

\end{abstract}


\clearpage

\section{Introduction}

Clusters assembled from metal particles are scientifically interesting
and technologically important \cite{Book}.
Physical phenomena in clusters of atoms and molecules do not 
necessarily have the same features as in  bulk phases.
Melting of metal clusters
was already studied  by computer simulations
\cite{Book,Wanda,Garzon,Manninen,Nielsen,Duxbury,Lewis}, and in experiments
\cite{Book,Buffat,Martin,Exppb}.
In contrast, the studies of clusters
at even higher temperatures, i.e., approaching
the liquid-vapour transition are 
scarce. Exceptions are early Monte Carlo and 
Molecular Dynamics (MD) simulations for very small Lennard-Jones clusters 
that extended to the temperature region of the vapour phase 
\cite{Abraham,Etters,Kaelberer,Rao}.
Related simulations were also performed in the field of nuclear fragmentation
\cite{Pandharipande}, and recently in the study of fragmentation 
for argon clusters \cite{Vladimir}. 
Understanding the behavior of nanoparticles at such high 
temperatures may also shed light onto the process of
achieving plasma and controlled fusion in a gas of clusters \cite{Ditmire}.

Melting of 
lead clusters of sizes ranging from a few to about $100$ nm 
were investigated by high-sensitivity optical reflectance and dark-field
electron microscopy \cite{Exppb}. 
Spherical and non-spherical nanometric
Pb inclusions in inert amorphous matrices of $SiO$ and $Al_2O_3$
were studied and it was found that melting  
is enhanced by surface and 
curvature effects.  These studies have been carried out up to the
bulk melting point of lead: $T_{sl}=600.56$ K, while 
the liquid-vapour transition temperature for bulk lead is: $T_{lv}=2033.16$ K.

In this letter we present a MD simulation study for liquid magic number
Pb clusters containing 147, 309, 561,
and 1451 atoms. The high-temperature structures,
total energies, the mean-square displacements, and diffusion of atoms 
are analyzed. A calculation of shape oscillations for lead nanodroplets
in the  liquid drop model has been carried out.

\section{Method and computational details}

Properties of metals in bulk, surfaces and clusters
are unlikely to be described well within the MD
method without a many-body potential \cite{Daw}.
We have used such a classical many-body potential for lead
\cite{Lim,Ong,Lead}.         
The set of parameters 
in the potential was determined by fitting to several measured 
properties of metal \cite{Furio}.
This potential was already used in MD simulations 
of the structural and vibrational
properties of lead clusters \cite{Lim,Ong}.
The bulk melting point  for the potential is $T_{sl}=618 \pm 4$ K \cite{Lead}.
This value was precisely determined by simulating coexisting
liquid and solid phases under constant energy.
For the present investigation of the high-temperature
behaviour of nanoparticles we have estimated the bulk liquid-vapour
transition temperature as $T_{lv}\sim  2050$ K.
The precise calculation of the liquid-vapour transition temperature for
a given potential
is in principle and computationally more involved (in comparison with the
calculation of the bulk melting point) and will be discussed 
elsewhere.

In this simulation
lead atoms were at $T=0$ K distributed as regular Mackay icosaheadra 
\cite{Mackay} of $147$, $309$, $561$, and $1415$ particles.
In reference \cite{Lim}, the authors find using the same potential that
at $T=0$ face-centered cubic morphologies are favoured over icosahedra.
This is the result of the low surface energy anisotropy between Pb(111)
and Pb(100), and of the high tensile surface stress which inhibits icosahedral
structure. However, at $T>0$ the stress decreases and icosahedra become more 
favoured. In experiments with argon clusters and in simulations for clusters 
described by pair potentials (such as Lennard-Jones and Morse), there is now
clear evidence that for less than $1500$ atoms icosahedra are more stable
than fcc structures \cite{Xie,Wales}. The situation is less clear for metallic
clusters. The majority of experimental and theoretical studies find that icosahedral 
clusters are also energetically preferred for metallic clusters of less than
$1500$ atoms, even at $T=0$ \cite{Clevland,Manninen}. Actually,
the morphology of the energetically preferred structure for solid clusters 
at low temperature is not important for our study of liquid nanoparticles at
high temperatures. Due to our wish to compare the results of simulations with
future experimental studies of melting and evaporation we have chosen 
icosahedra as starting point in our simulation.  
As in other MD simulations of metal clusters 
\cite{Wanda,Garzon,Manninen,Nielsen,Duxbury,Lewis} 
no boundary conditions have been employed.  
A time step of $7.33 \times 10^{-14}$ s was used. 
MD simulations with damped dynamics were first performed to extract the kinetic
energy until the clusters reached the structure corresponding to 
a local energy minimum. Then the clusters were heated  
up to temperatures ($T < 2000$ K)
where evaporation of atoms starts
within a typical time scale of classical MD simulations.  
The embedded atom potentials are known to show realistic evaporation, i.e., 
to be much less prone to enhanced evaporation
during MD simulation in comparison with the pair potentials. 
First the ``heating'' runs at constant temperature  were carried out
for $5\times10^4$ time steps. 
Because of the difficulties in achieving 
equilibrium in the liquid temperature region, rather long 
runs of $10^6$ time steps (i.e., $7.3$ ns) were used to
calculate the cluster properties.
The temperature was controlled by rescaling the particle velocities
at each time step.  
Note that the potential used here (and other embedded atom potentials)
are not applicable at extremely high temperatures, for example at 
$T > 2000$ K and approaching plasma conditions. 
These potentials are not fitted for 
this temperature region and the resulting low atomic coordination.

\section{Results and Discussion}

\subsection{Structures and mobility}

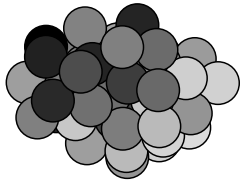
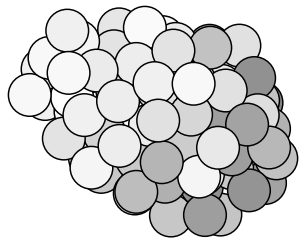
\begin{figure}
\caption{Atomic positions showing:
(a) non-spherical shape of the liquid cluster of 147 atoms at $T=1300$ K
after $6.96$ ns of time evolution,
(b) the same sample after $7.32$ ns.
The relative darkness of an atom is proportional to its square displacement 
during the run. Black particles are those with the maximum of mobility.}
\label{fig1}
\end{figure}

The equilibrium shape of a macroscopic
liquid metal drop is a sphere because of the negligible
anisotropy of the liquid-vapour interface free energy in comparison 
with the one of the solid-vapour interface.
We have found that on the simulation time scale the
liquid clusters exhibit substantial shape fluctuations at high temperatures.  
This tendency towards transient formation of non-spherical shapes
decreases with increasing cluster size, but it is still present even for the 
largest cluster of $1415$ atoms studied here. We may mention that 
non-spherical shapes were also found in experimental studies of nanometric
lead inclusions \cite{Exppb}.  
A snapshot of atomic positions producing a 
very elongated lead droplet is presented in Fig.~\ref{fig1}  (a).
In the further time evolution this cluster separates into 
two smaller droplets as shown in Fig.~\ref{fig1}  (b).
The colour of the atoms reveals that the cluster emits a small subcluster. 
The process presented in Fig.~\ref{fig1}  is the only occurrence we found 
(within our long, but limited simulation time) for a droplet 
separation into two parts, although
several other extremely elongated shapes
similar to Fig. ~\ref{fig1} (a) were observed, especially for the smallest 
cluster and above
$T>1000$ K. For most liquid droplets 
a typical process at these temperatures is single atom evaporation.
A similar tendency of less intense evaporation for small as compared to
large nanoparticles was recently found in a MD study of melting of
argon clusters \cite{Finland}.
Another rare event -  a dimer evaporation was also detected.
We have found that the number of evaporated atoms strongly
depends on the length of time evolution during the MD simulation.
For example, a liquid nanoparticle of $1415$ atoms at $T=1600$ K
completely vaporizes after $3\times 10^5$ time steps, whereas after 
$10^5$ time steps only 4 atoms evaporate.
For an almost immediate (within $\sim 10$ ps) complete ``explosive''
disintegration of lead 
nanoparticles of these sizes much higher temperatures of the order of $5000$ K
are needed. However, as pointed out above,
the reliability of the potential used is 
questionable in this temperature region.

In most bulk liquids the coordination number is $\sim 8-11$, i.e., it is 
smaller than the coordination number of,
for example,  $12$ which is the property of
close-packed bulk solid fcc metals \cite{Croxton}.
In addition, as a result of surface effects, the average coordination 
number in clusters is always smaller than in the bulk.
The average coordination numbers for all liquid nanoparticles studied here at
several temperatures are given in Table \ref{table1}.
Our calculation of the coordination numbers
(and other cluster properties which we present below) extends 
to the beginning of the evaporation regime, i.e., up to 
$T=1200$ K. 
We found that at $T=1000$ K and $T=1200$ K, after several nanoseconds,
one or two atoms evaporated from the liquid clusters of $309$, $561$, and
$1415$ atoms.
After such an evaporation process,  
the calculations were continued  only for the remaining liquid cluster. 
From Table \ref{table1} one can see how the average 
coordination number roughly increases with cluster size 
and decreases with temperature.

Steps in the caloric curve are a signature of melting 
\cite{Book,Wanda,Garzon,Manninen,Nielsen,Duxbury,Lewis}.
We also studied the potential energy per atom, $E$,
as a function of temperature from  
melting up to the evaporation regime. It was found that
the $E=f(T)$ function is linear for all studied nanoparticles
in that temperature region.

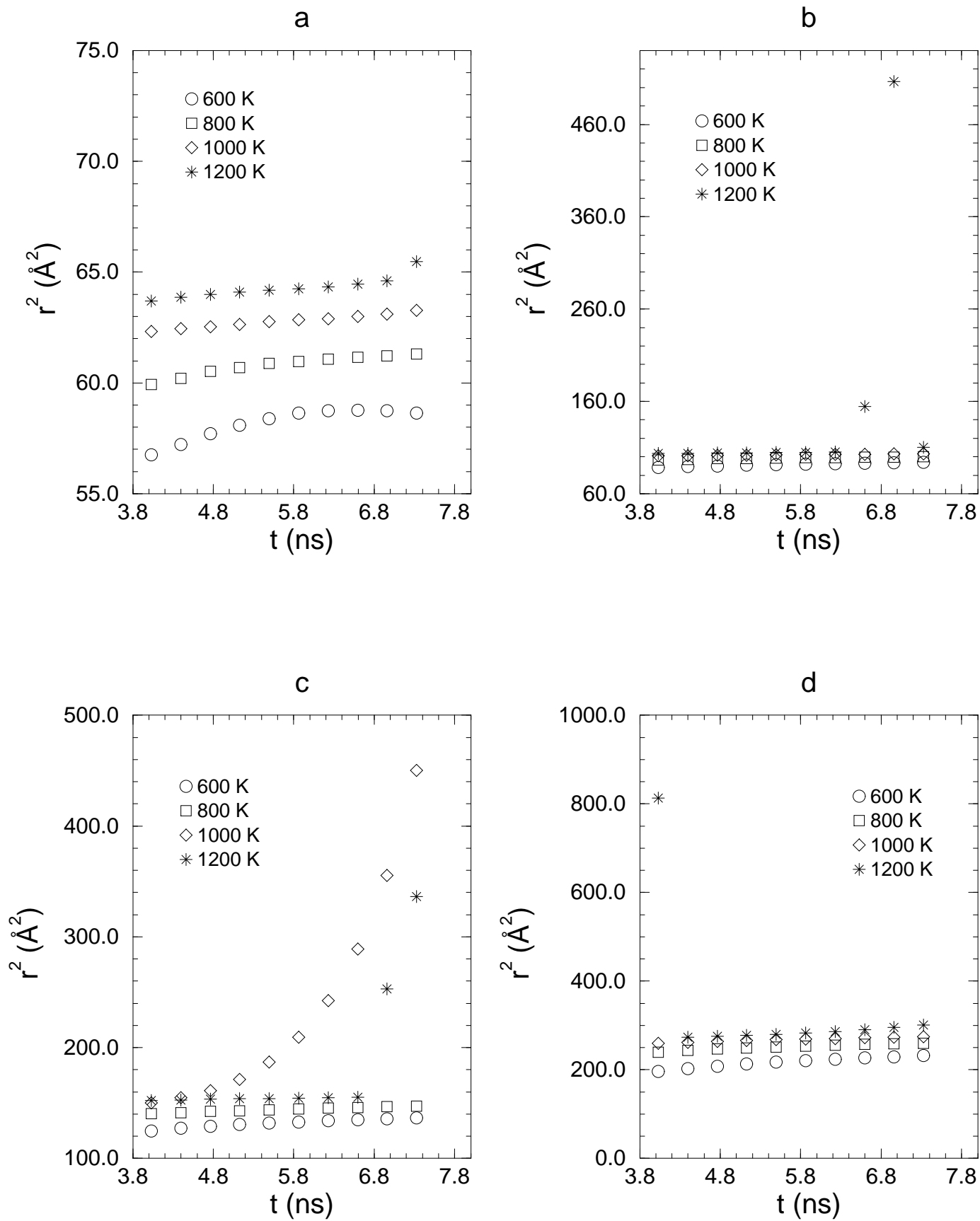
\begin{figure}
\caption{
Mean-square displacements  for liquid lead nanoparticles
at several temperatures: (a) 147, (b) 309, (c) 561, (d) 1415 particles.}
\label{fig2}
\end{figure}

Atomic configurations presented in Fig. ~\ref{fig1} 
show the mobility of the atoms on a  qualitative level. 
The mean-square displacements for the clusters after several nanoseconds
of simulation and 
at several temperatures are shown in Fig.~\ref{fig2}.
At lower temperatures ( $T < 1000$ K), 
the mean-square displacements rise approximately linearly with time.
Such a behaviour is typical for liquid diffusion. 
Sharp increases 
at higher temperatures are the results of evaporation, or the onset of
evaporation. 
As pointed out above, the atom which has evaporated (and moved from the
surface of the nanodroplet at a distance larger than the cutoff radius of
the potential, i.e., $> 5.6$ \AA) is excluded from further consideration.
Note that as a result of evaporation from the cluster of $561$
atoms the mean-square displacements are bigger at $1000$ K (560 atoms
in the liquid cluster), than at $1200$ K (559 atoms).

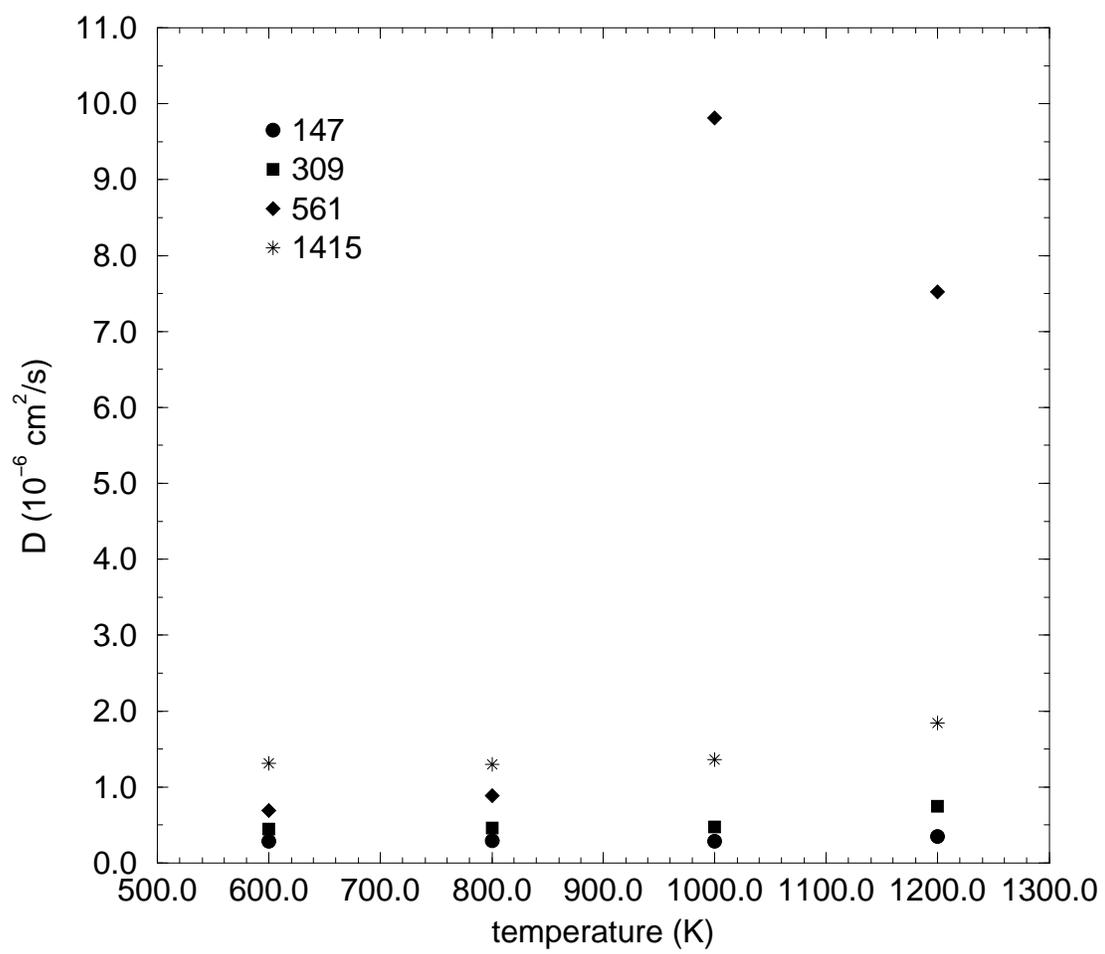
\begin{figure}
\caption{
Diffusion coefficients for liquid lead clusters as a function 
of the temperature.}
\label{fig3}
\end{figure}

The diffusion coefficients of bulk solid  
and bulk liquid phases are orders of magnitude
different ($10^{-9}$ $cm^2 s^{-1}$ vs $10^{-5}$ $cm^2 s^{-1}$).
The temperature dependence of the diffusion coefficient of lead clusters 
is shown 
in Fig. ~\ref{fig3}. The values are typical for diffusion in the liquid phase. 
As for the mean-square displacements, a jump of the  diffusion coefficient
is the result of the onset of evaporation.

\subsection{Shape oscillations of liquid lead clusters}

It is interesting to estimate the influence of the size of a lead 
nanodroplet on its shape oscillations. Many phenomena 
in cluster physics can be analyzed using the tools developed in 
nuclear physics \cite{Schmidt}. 
For nuclei some phenomena, for example magic numbers, can only be 
studied within independent particle models. Other phenomena, for example
binding energies, vibrational properties, and fission of heavy nuclei, 
are studied using collective (continuum) models, such as the liquid drop 
model. Experimental and theoretical investigations have shown that the 
same two types of models can be applied for metallic clusters \cite{Schmidt}.
The fission measurements for small gold clusters were interpreted using the
nuclear liquid drop model and the results indicated that the vibrational 
properties of metallic clusters, nuclei, and liquid drops are similar
\cite{Saunders}. A common approach to the analysis of vibrational properties
of clusters is either by direct diagonalization of the dynamical matrix, or by
calculating the Fourier transform of auto-correlation functions 
in a MD simulation. For liquid clusters bigger than $100$ atoms as studied in 
this work, these traditional approaches give a broad, structureless line in 
the frequency spectrum \cite{Kristensen,Ong}. 
Instead, we analyze the shape oscillations of liquid lead
nanoparticles using the nuclear liquid drop model \cite{Bohr}.
In this model surface and volume modes are analyzed separately and the results
are expressed by macroscopic properties of the metal. 
Three kinds of shape oscillations are possible: surface oscillations, 
modes of vibration involving compression, and coupled oscillations.

The distance of the surface 
of a slightly deformed liquid drop from the origin,
$R(\theta,\phi)$, can be expanded as
\begin{equation}
R(\theta,\phi) = R_0 \left (1 + \sum_{\lambda\mu} \alpha_{\lambda\mu} 
Y^{*}_{\lambda\mu}(\theta,\phi)\right),
\label{eq:1}
\end{equation}
where $R_0$ is the equilibrium radius, $\alpha_{\lambda\mu}$ are 
normal coordinates, and $Y_{\lambda\mu}(\theta,\phi)$ are the spherical
harmonics \cite{Bohr}. 
The equations of oscillatory motion are 
\begin{equation}
{\ddot {\alpha}}_{\lambda\mu} + \omega(s)^2_{\lambda}\alpha_{\lambda\mu} = 0.
\label{eq:2}
\end{equation}
Eigenfrequencies of the surface oscillations are given by
\begin{equation}
\omega(s)_{\lambda}^2 = \frac {\sigma \lambda (\lambda -1)(\lambda+2)}
{\rho_0 R_0^3}, 
\label{eq:3}
\end{equation}
where $\sigma$ is the surface tension and $\rho_0$ is the density \cite{Bohr}.
In Eqs. (1)-(3), the value $\lambda=0$ represents a compression without
change of shape, whereas $\lambda=1$ describes a translation of a droplet
as whole. This can be seen by considering Eq. (1) and respectively,
the volume of an incompressible drop and a small translation of 
the center of mass along the axis $\theta=0$. Therefore, the lowest order
surface oscillations are quadrupole modes $\lambda=2$.

Compressional oscillations in liquid droplets also exist. Small deviations of
the density $\rho(\vec{r})$ 
which are solutions of the linearized hydrodynamical 
equations that lead to the wave equation are given by
\begin{equation}
\delta \rho (\vec{r},t) = \rho_0 j_{\lambda} (k_{n\lambda}r) Y^*_{\lambda\mu} 
(\theta,\phi)\alpha_{n\lambda\mu}(t),
\label{eq:4}
\end{equation}
where $j_{\lambda}$ is a spherical Bessel function. 
If the boundary condition 
is such that $\delta \rho(\vec{r},t)$ vanishes at $r=R_0$, then the eigenvalue 
equation is 
\begin{equation}
j_{\lambda}(k_{n\lambda}R_0)=0.
\label{eq:5}
\end{equation}
The compressional eigenfrequencies are given by 
\begin{equation}
\omega(c)_{n\lambda}= u_c k_{n\lambda},
\label{eq:6}
\end{equation}
where $u_c$ is the sound velocity in the liquid.

The ratio of the eigenfrequencies of surface and compressional modes (Eqs. (3)
and (6)) slowly decreases as the number of atoms in the cluster increases.
In general, if the liquid cluster (or nucleus) is small, 
it is important to consider coupling of the  surface and 
compressional modes. These coupled modes are solutions of the wave 
equation with a modified boundary condition such that the pressure
induced by the shape oscillations is in equilibrium with the pressure induced
by the compressional oscillations \cite{Bohr}. 
The corresponding eigenvalue equation of the
coupled modes is 
\begin{equation}
\frac {1} {j_{\lambda}(k_{n\lambda}R_0)} 
\left [\frac {\partial} {\partial r} j_{\lambda}(k_{n\lambda}r)\right ]
_{r=R_0} =
\frac {\lambda} {R_0} \frac {\omega (c)^2_{n\lambda}} {\omega(s)^2_{\lambda}}.
\label{eq:7}
\end{equation}

The surface, compressional, and coupled
frequencies were calculated using the 
experimental data for bulk lead: $u_s = 1800$ $m s^{-1}$ \cite{Kaye},
$\rho_0 = 10678$ $kg m^{-3}$ \cite{Www}, and $\sigma=0.453$ $N m^{-1}$
\cite{Kaye}. The average radii of approximately spherical clusters were taken
from configurations obtained in MD simulations at $600$ K. 
The calculation shows that for these lead nanodroplets the frequencies of
coupled modes are approximately equal to the frequencies of surface modes.
The size dependence of the surface and compressional frequencies is presented
in Fig. 4. 
Frequencies of compressional modes are of course
bigger than the corresponding frequencies of surface modes. 
The frequencies increase with increasing value of $\lambda$ and  
with decreasing size of the nanodroplets.
This agrees with our results of MD simulations where the most pronounced shape
oscillations were found for the smallest liquid cluster of $147$ atoms.

These  results provide for the first time an insight into the behavior 
of neutral liquid metal clusters at high temperatures,
up to the beginning of the evaporation regime. 
Experimental investigations of clusters in this temperature range,  
either using optical,
diffraction techniques, and  mass spectroscopy \cite{Book,Buffat,Martin,Exppb},
or cluster fragmentation properties 
\cite{Vladimir,Ditmire,Udo} are desirable.

\acknowledgments

This work has been supported by the Volkswagen Foundation and the
Deutsche Forschungsgemeinschaft.

\clearpage

\begin{table}
\caption{
The average coordination numbers for liquid lead clusters
of a given total number $N$ of particles, calculated after $7.3$ ns 
of simulation.
If evaporated atoms exist their number is given in parentheses;
they were 
excluded in the calculation of the average coordination number and other
cluster properties.}
\label{table1} 
\begin{tabular}{l l l l l} 
Temperature (K)  & $N=147$ & $N=309$ & $N=561$  & $N=1415$ \\ 
\hline 

$600$  & $8.99$ & $9.00$ & $9.34$ & $9.72$\\ 

$800$  & $8.75$ & $8.84$ & $8.93$ & $9.34$\\

$1000$  & $8.44$ & $8.43$ & $8.94 (1)$ & $9.04$ \\ 

$1200$   & $8.41$ & $8.13 (1)$ & $8.64 (2)$ & $8.88 (2)$ \\ 
\end{tabular}
\end{table}

\clearpage

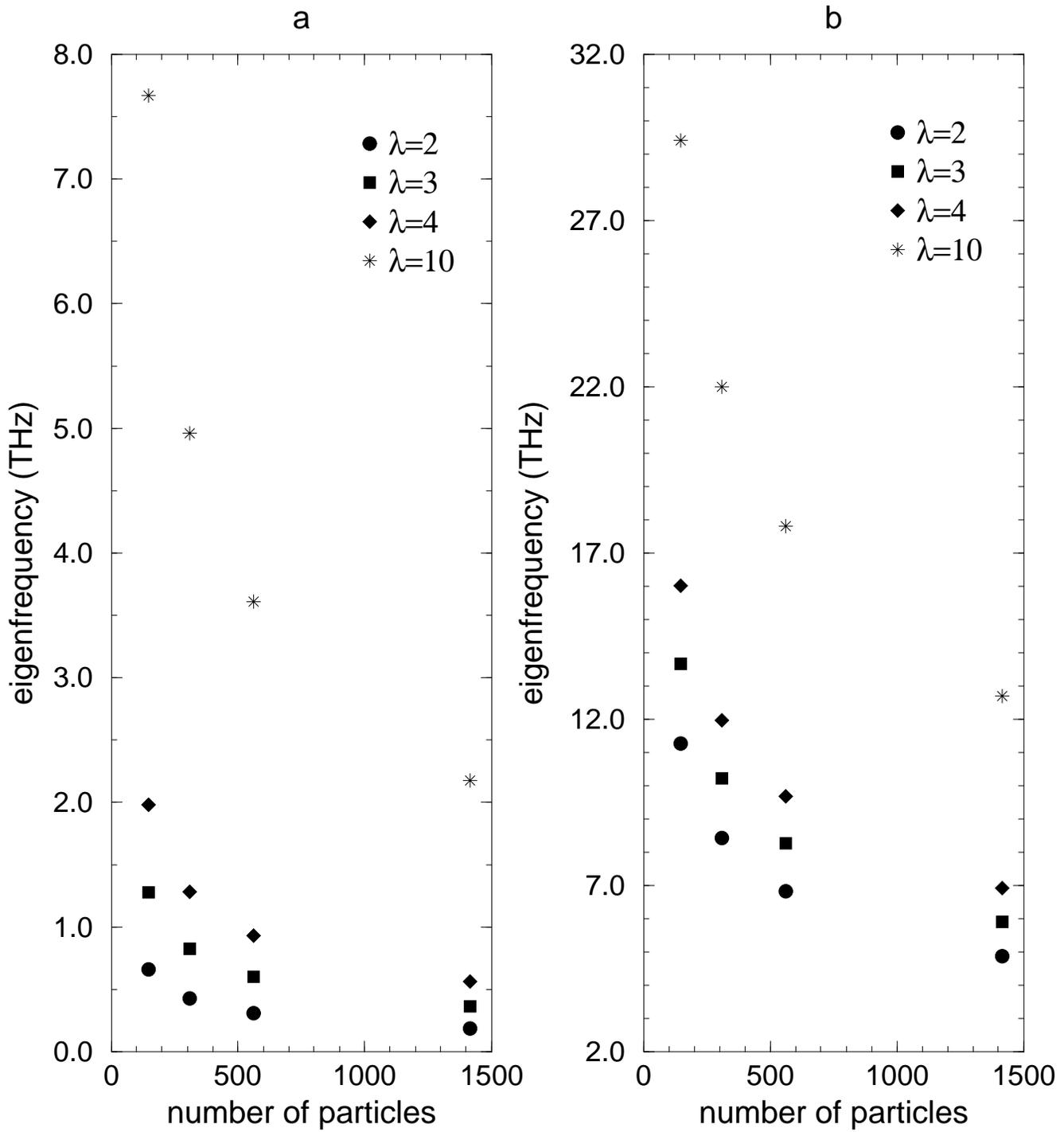
\begin{figure}
\caption{
Eigenfrequencies of lead nanodroplets in the liquid drop model
as a function size: 
(a) surface oscillations, 
(b) compressional oscillations ($n=1$).}
\label{fig4}
\end{figure}


\clearpage



\begin{references}


\bibitem{Book}
Large Clusters of Atoms and Molecules, ed. T. P. Martin
(Kluwer, Dordrecht, 1996).

\bibitem{Wanda}
F. Ercolessi, W. Andreoni, and E. Tosatti,
Phys. Rev. Lett. {\bf 66} (1991) 911.

\bibitem{Garzon}
I. L. Garzon and J. Jellinek,
Z. Phys. D {\bf 20} (1991) 235.

\bibitem{Manninen}
S. Valkealahti and M. Manninen,
Phys. Rev. B {\bf 45} (1992) 9549. 

\bibitem{Nielsen}
O. H. Nielsen, J. P. Sethna, P. Stoltze, K. W. Jacobsen, and 
J. K. Norskov. 
Europhys. Lett. {\bf 26} (1994) 51.

\bibitem{Duxbury}
X. Yu and P. M. Duxbury,
Phys. Rev. B {\bf 52} (1995) 2102.

\bibitem{Lewis}
L. J. Lewis, P. Jensen, and J. L. Barrat,
Phys. Rev. B {\bf 56} (1997) 2248.


\bibitem{Buffat}
P. Buffat and J. P. Borrel,
Phys. Rev. A {\bf 13} (1976) 2287.

\bibitem{Martin}
T. P. Martin, U. N{\" a}her, H. Schaber, and U. Zimmermann,
J. Chem. Phys. {\bf 100} (1994) 2322.

\bibitem{Exppb} 
R. Kofman, P. Cheyssac, A. Aouaj, Y. Lereah, G. Deutscher, P. Cheyssac, 
T. Ben-David, J. M. Penisson, and A. Bourret,
Surf. Sci. {\bf 303} (1994) 231. 


\bibitem{Abraham}
J. K. Lee, J. A. Barker, and F. F. Abraham,
J. Chem. Phys. {\bf 58} (1973) 3166.

\bibitem{Etters}
R. D. Etters and J. Kaelberer,
Phys. Rev. A {\bf 11} (1975) 1068.

\bibitem{Kaelberer}
J. Kaelberer and R. D. Etters,
J. Chem. Phys. {\bf 66} (1977) 3233.

\bibitem{Rao}
M. Rao, B. J. Berne, and M. H. Kalos,
J. Chem. Phys. {\bf 68} (1978) 1325. 

\bibitem{Pandharipande}
A. Vicentini, G. Jacucci, and V. R. Pandharipande,
Phys. Rev. C {\bf 31} (1985) 1783.


\bibitem{Vladimir}
V. N. Kondratyev, H. O. Lutz, and S. Ayik,
J. Chem. Phys. {\bf 106} (1997) 7766.


\bibitem{Ditmire}
T. Ditmire, J. W. G. Tisch, E. Springate, M. B. Mason, N. Hay, R. A. Smith,
J. P. Marangos, and M. H. R. Hutchinson, 
Nature {\bf 386} (1997) 54.



\bibitem{Daw}
M. S. Daw, S. M. Foiles, and M. I. Baskes,
Mater. Sci. Rep. {\bf 9} (1993) 251. 



\bibitem{Lim}
H. S. Lim, C. K. Ong, and F. Ercolessi, 
Surf. Sci. {\bf 269/270} (1992) 1109.

\bibitem{Ong}
H. S. Lim, C. K. Ong, and F. Ercolessi, 
Z. Phys. D {\bf 26} (1993) S45.

\bibitem{Lead}
G. Bilalbegovi{\' c}, F. Ercolessi, and E. Tosatti,
Europhys. Lett. {\bf 17} (1992) 333.

\bibitem{Furio}
F. Ercolessi, M. Parrinello, and E. Tosatti,
Philos. Mag. A {\bf 58} (1988) 333.

\bibitem{Mackay}
A. L. Mackay, Acta Crystallogr. {\bf 15} (1962) 916.

\bibitem{Xie}
J. Xie, J. A. Northby, D. L. Freeman, and J. D. Doll,
J. Chem. Phys. {\bf 91} (1989) 612.


\bibitem{Wales}
J. P. K. Doye and D. J. Wales,
Z. Phys. D {\bf 40} (1997) 466. 

\bibitem{Clevland}
C. L. Clevland and U. Landman,
J. Chem. Phys. {\bf 94} (1991) 7376.

\bibitem{Finland}
A. Rytk{\" o}nen,  S. Valkealahti, and M. Manninen,
J. Chem. Phys. {\bf 106} (1997) 1888. 


\bibitem{Croxton}
C. A. Croxton, Statistical Mechanics of the Liquid Surface
(John Wiley, New York, 1980).



\bibitem{Schmidt}
Nuclear Physics Concepts in the Study of Atomic Cluster Physics,
eds. R. Schmidt, H. O. Lutz, and R. Dreizler
(Springer, Berlin, 1992).


\bibitem{Saunders}
W. A. Saunders, 
Phys. Rev. Lett. {\bf 64} (1990) 3046 and references therein. 


\bibitem{Kristensen}
W. D. Kristensen, E. J. Jensen, and R. M. J. Cotterill, 
J. Chem. Phys. {\bf 60} (1974) 4161.


\bibitem{Bohr}
A. Bohr and B. R. Mottelson, Nuclear Structure, Vol. II
(Benjamin, Reading, 1975).



\bibitem{Kaye}
G. W. C. Kaye and T. H. Laby, 
Tables of physical and chemical constants, 
(Longman, London, 1973).


\bibitem{Www}
WebElements, http://www.shef.ac.uk/{$\sim $}chem/web-elements/,
(University of Sheffield, Sheffield, 1997).


\bibitem{Udo}
U. Werner, K. Beckord, J. Becker, and H. O. Lutz,
Phys. Rev. Lett. {\bf 74} (1995) 1962.





\end{references}
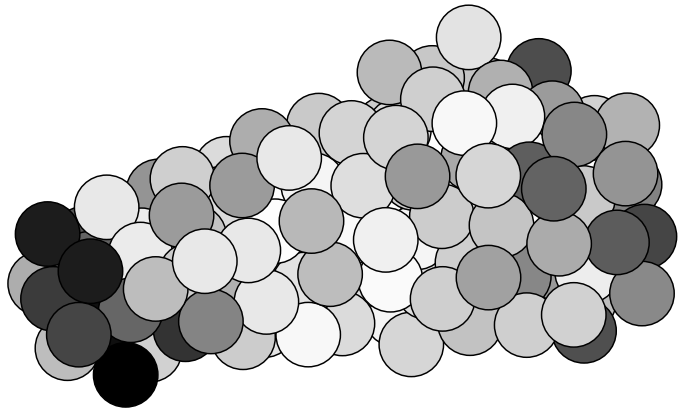
\end{document}